\documentclass[aps,floatfix]{revtex4-2}

\usepackage{amssymb,hyperref,hypcap,
ae,tikz,color}
\usepackage{latexsym,textcomp,slashed,epsfig}
\usepackage[T1]{fontenc}
\usepackage{verbatim}

\usepackage{graphicx}
\usepackage{dcolumn}
\usepackage{bm}
\usepackage{amsmath}
\usepackage{graphicx}
\usepackage{listings}
\usepackage[T1]{fontenc}
\usepackage{xcolor}
\usepackage{lmodern}
\usepackage{listings}
\usepackage[caption=false]{subfig}
\usepackage{color}
\usepackage{mathtools}

\hypersetup{
	colorlinks=true,
	citecolor=green,
    linkcolor=magenta,
	}

\begin{document}

\title{{\bf An investigation on gravitational entropy of cosmological models}}

\author{Samarjit Chakraborty$^{1}$$^{\dag}$}
\author{Sarbari Guha$^{1}$$^{\S}$}
\author{Rituparno Goswami$^{2}$$^{\P}$}
\affiliation{ $^1$Department of Physics, St.Xavier's College (Autonomous), Kolkata 700016, India \\ $^2$Astrophysics and Cosmology Research Unit, School of Mathematics, Statistics and Computer Science, \\ University of KwaZulu-Natal, Private Bag X54001, Durban 4000, South Africa\\
$^{\dag}$samarjitxxx@gmail.com \\ $^{\S}$srayguha@yahoo.com; guha@sxccal.edu \\ $^{\P}$goswami@ukzn.ac.za}

\maketitle
\section*{Abstract}
In this paper we investigate the entropy of the free gravitational field for a given epoch for some well known isotropic and anisotropic cosmologies. We use the definition of gravitational entropy proposed by Clifton, Ellis and Tavakol, where the 2-index square root of the 4-index Bel-Robinson tensor is taken to be the energy momentum tensor for the free gravity. We examine whether in the vicinity of the initial singularity, the ratio of energy density of free gravity to that of matter density goes to zero, validating Penrose conjecture on Weyl curvature. Whenever this is true, the gravitational entropy increases monotonically with time, leading to structure formation. For the models considered by us, we identify the conditions for which the Weyl curvature hypothesis is valid, and the assumptions under which it is validated, or otherwise.
\bigskip

KEYWORDS: General Relativity; Gravitational entropy; Weyl curvature hypothesis; Cosmological models

\section{Introduction}

The proposal of gravitational entropy attempts to provide a sense of sequence to gravitational processes, and remains as one of the open problems in General Relativity (GR) till today. Although a suitable definition of gravitational entropy in the case of stationary black holes was available in literature for quite some time \cite{bh}, but a universally agreeable analogue in the case of cosmology has been under the process of formulation till late.

It is well-known that GR is plagued by the problem of spacetime singularities. However, Roger Penrose \cite{Penrose} put forward his belief that this problem was a consequence of the limitations of the ``very notion of spacetime geometry'' and the corresponding physical laws. According to him, the problem of spacetime singularities held the key to the ``origin of the \emph{arrow of time}''. Several researchers \cite{arrow,PL} based their studies on the notion of this arrow of time.

Originally Penrose proposed the \textit{Weyl curvature hypothesis} (WCH), which merely required that the Weyl tensor should be zero at the big bang singularity \cite{Rothman}, and the subsequent evolution of the universe must be close to the homogeneous and isotropic Friedman-Lemaitre-Robertson-Walker (FLRW) model. Assuming the existence of ``gravitational entropy'', Penrose argued that \emph{the principle of increase
of entropy} implies that the big bang singularity should be of low entropy, which is related to the ``absence of clumping of matter'', and hence to the absence of Weyl tensor, so that the big bang singularity must have been a regular one. Thus the universe could evolve to its observed FLRW form and be consistent with the second law of thermodynamics \cite{PL,PC}. If matter was approximately under thermodynamic equilibrium during the big bang, then it requires a corresponding low entropy of the gravitational field. Although the big bang would normally be considered as a state of maximum entropy (assuming thermal equilibrium), but in reality the entropy of the universe is increasing. This apparent paradox may be due to the omission of gravitational degrees of freedom, and the gravitational entropy of the big bang was actually low. Penrose \cite{Penrose} argued that gravitational entropy must be defined from the free gravitational field. An increase in gravitational entropy would imply an increase in the local anisotropy (thereby facilitating structure formation), which can be quantified by the shear tensor. By analyzing the trace free Bianchi identities we can also suggest that this shear tensor affects the evolution of the Weyl tensor \cite{MB}, thereby establishing a physical relationship between gravitational entropy and the Weyl tensor. Although a conformally flat perfect fluid spacetime has vanishing shear and vorticity, and the metric is of FLRW type \cite{ST}, but conformally flat spacetimes with diagonal trace-free anisotropic pressure and zero cosmological constant is found to possess a simple equation of state where the energy per particle density depends only on the shear scalar \cite{BR}. This provides us with one more reason to connect local anisotropy i.e. gravitational entropy to the Weyl tensor.

Penrose proposed that the gravitational entropy should be related to a suitable measure of the Weyl curvature, and the condition of low entropy should enforce constraints on the Weyl curvature. All these implied that some suitable dimensionless scalar must be asymptotically zero. Therefore, the determination of the gravitational entropy function requires the construction of this scalar function.

In 1982, Goode and Wainwright \cite{GW1982} presented a new formulation of the two classes of Szekeres solutions of the Einstein field equations, and provided a general analysis of the scalar polynomial curvature singularities of these solutions, and of their time-evolution. They identified the solutions
which are close to an FLRW model near the initial singularity, or in the later stages of evolution.

Wainwright and Anderson  discussed the evolution of a class of exact spatially homogeneous cosmological models of Bianchi type $VI_{h}$ \cite{wainander}. It is known that solutions of type $VI_{h}$ cannot approach isotropy asymptotically at large times. They infact  become asymptotic to an anisotropic vacuum plane wave solution. Nevertheless, for these solutions the initial anisotropy decays and  leads to a stage of finite duration in which the model is close to isotropy. Depending on the choice of parameters in the solution, this quasi-isotropic stage can commence at the initial singularity, in which case the singularity is of the type known as ``isotropic'' or ``Friedmann-like''. The existence of this quasi-isotropic stage implies that these models can be compatible in principle with the observed universe. Inspired by the WCH, Goode and Wainwright \cite{GW} gave the geometric definition of the concept of `isotropic singularity' and showed that the Weyl tensor is dominated by the Ricci tensor at this scalar polynomial curvature singularity. Husain \cite{Husain} examined the WCH in the context of the Gowdy cosmology. He calculated the expectation values of the square of the Weyl curvature in states of clumped and unclumped gravitons and found that the curvature contains information about the gravitational entropy.

Senovilla \cite{Senovilla} showed that the Bel-Robinson tensor is quite suitable for providing a measure of the energy of gravitational fields. Following his work, there were several attempts to define the gravitational entropy on the basis of the Bel-Robinson tensor on one hand, and also in terms of the Riemann tensor and its covariant derivatives \cite{PL,PC}. Lake and Pelavas \cite{PL} introduced a class of ``gravitational epoch'' functions which were dimensionless scalars, one of which was built from the Riemann tensor and its covariant derivatives only, denoted by $P$. Other alternative functions involving the Bel-Robinson tensor were also suggested by them. They analyzed whether such functions could be regarded as gravitational entropy function or not. Other dimensionless scalars have also been considered, for example by \cite{GCW}, which are constructed from the Riemann tensor and its covariant derivatives.

In spite of all these efforts, there was still doubt regarding the definition of gravitational entropy in a way analogous to the thermodynamic entropy, which would be applicable to all gravitational systems. Attempts were also made to explain the gravitational entropy of black holes. Among them one interesting approach is to handle the problem from a phenomenological point of view as proposed in \cite{entropy1} and expanded in \cite{entropy2}, for the purpose of testing the WCH against the expressions for the entropy of cosmological models and black holes. They considered a measure of gravitational entropy in terms of a scalar derived from the contraction of the Weyl tensor and the Riemann tensor, and matched it with the Bekenstein-Hawking entropy \cite{SWH1,Bekenstein}. Recently, Guha and Chakraborty \cite{GC1} investigated whether the prescriptions for calculating gravitational entropy as proposed in \cite{entropy1} and \cite{entropy2} could be applied to the case of the accelerating black holes. They found that such a definition of gravitational entropy works pretty well for the accelerating black holes and charged accelerating black holes, except for the rotating charged accelerating metric.

An important proposal was offered by Clifton et al. \cite{CET}, who provided a measure of gravitational entropy based on the square root of the Bel–Robinson tensor which was motivated by thermodynamic considerations, and has a natural interpretation as the effective super-energy-momentum tensor of free gravitational fields. They applied this construction to several cases, including cosmological ones, and found that the specific form of this measure depended on the nature of the gravitational field, namely, whether it was Coulomb-like or wave-like. However, this definition of gravitational entropy is only valid for General Relativity, where the Bel-Robinson tensor can be defined in this way. In the subsequent text, we refer to this formulation as the ``CET proposal''.

Bolejko \cite{cet5} showed that both the notion of gravitational entropy of the universe (associated with inhomogeneity) and the cosmic
no-hair conjecture (that a universe dominated by dark energy should asymptotically approach a homogeneous and isotropic de Sitter state) are simultaneously valid and are not contradictory. It was found that a universe with a positive cosmological constant and nonpositive spatial curvature in fact approaces the de Sitter state, but at the same time keeps generating the gravitational entropy.

In the paper \cite{CET}, the authors considered scalar perturbations of a FLRW geometry. They found that the gravitational entropy function behaved like the Hubble weighted anisotropy of the gravitational field, which therefore increases with structure formation. On the other hand, the FLRW metric is conformally flat \cite{Ellis}, and so the Bel–Robinson tensor has vanishing components. The gravitational epoch function $W$ generated by the Bel–Robinson tensor, is also vanishing, resulting in a vanishing gravitational entropy. In fact, gravitational entropy is identified with the presence of inhomogeneity, which requires both anisotropy and a non-zero $W$.

The intermediate homogenization of inhomogeneous cosmological models was studied in \cite{cet6} along with the problem of gravitational entropy. All definitions of entropy examined in this paper yielded decreasing gravitational entropy during the homogenization process, which implies that the gravitational entropy may actually decrease in some cases.

Gr{\o}n and Hervik \cite{gron} investigated the evolution of different measures of `gravitational entropy' in Bianchi type I and Lema\^{\i}tre–Tolman–Bondi (LTB) universe models. They found that the WCH remains secure if one considers the non-local version of the conjecture.
In their paper, Mishra and Singh \cite{cet7} investigated whether the inhomogeneous cosmological models could be motivated on the basis of thermodynamic grounds and a particular minimal void LTB inhomogeneous model was chosen for the analysis. They examined several definitions of gravitational entropy and found that the Weyl curvature entropy exhibits satisfactory thermodynamic behavior in the case of inhomogenous cosmologies.

Sussman \cite{cet0} introduced a weighed scalar average formalism (the `q-average' formalism) for the study of the theoretical properties and the dynamics of spherically symmetric LTB dust models and explored the application of this formalism to a definition of a gravitational entropy functional proposed by Hosoya et al (HB proposal) \cite{HB}. Subsequently, Sussman and Larena \cite{cet1} considered the generic LTB dust models to probe the CET proposal and the HB proposal, along with a variant of the HB proposal, suggesting that the notion of gravitational entropy is a theoretically robust concept which can also be applied to other general spacetimes.

The evolution of the CET gravitational entropy was also studied by the same authors in \cite{cet2} for the local expanding cosmic CDM voids using a non-perturbative approach. Marozzi et al \cite{cet4} calculated the gravitational entropy of the large scale structures of the universe in the linear regime, where it can be described by the perturbed FLRW spacetime. This entropy arises from the averaging made over an extended region and explains the formation of large-scale structure in the Universe. The results obtained in \cite{cet2} for the gravitational entropy agreed well with their results, when the LTB evolution is in its linear regime, thus providing us with a connection between the local physics and the large scale linear regime.

In \cite{cet3} it was pointed out that the formation of numerous astronomical objects like the supermassive black holes (quasars), super-novae and dust, and the occurrence of several phenomenona like gamma ray bursts in the early universe are contradictory to the conventional mechanisms of its possible origin. The $\Lambda$CDM cosmology fails to explain several observations like the absence of central cusps with $\rho \sim r^{-1}$ in the dark matter distribution \cite{Primack}, presence of too many bright satellite galaxies at high $z$ \cite{list1,list2,list3}, or the larger value of observed angular momentum of galaxies. Although the standard cosmological model successfully describes the gross properties of the universe, yet fails in terms of several smaller details, both in the early universe at redshifts $z \sim 10$ and in the present time. The early universe is abundantly populated by quasars (\cite{quasar} and references therein), but it is practically impossible to create so many quasars in the young universe assuming the standard mechanism of BH formation by the process of matter accretion. These discrepancies may be removed if there exists a dark matter particle having life-time greater than the age of the universe at the time of recombination \cite{BDT}. Such observations suggest the existence of New physics beyond the standard theory, which therefore requires suitable modification (see also the references in \cite{cet3}).

A possible solution may lie in a different cosmological expansion law as indicated in \cite{DHT} and \cite{MM}. The supersymmetric Grand Unified models \cite{cet3,AD} consider the action of a scalar field, $\chi$, with non-zero baryonic number, $B$, due to which bubbles may be generated. Initially (after inflation) $\chi$ was away from the origin and when inflation is over it starts to evolve down to the equilibrium point, $\chi = 0$. Because of the inflationary expansion, the bubbles could become astrophysically large. Immediately after the formation of bubbles with large value of $\chi$, inhomogeneities developed in the energy density due to different equations of state in the regions inside and outside these bubbles. The big bang nucleosynthesis (BBN) inside or in the vicinity of the high-$B$ bubbles creates heavy elements more efficiently than that predicted in the standard model. This may lead to the observed distribution of the celestial bodies and a lot of dust at $z \sim 10$ \cite{cet3}. Also recently astronomers have shown a very strong possibility of an anisotropic universe unlike the standard assumption of a homogeneous and isotropic universe \cite{ani1,ani2}.

These informations provide a very strong motivation for us to explore the inhomogeneous and anisotropic cosmologies. Such models have also been studied recently \cite{ani3} for the resolution of the discrepancy between the Hubble parameter measured locally as opposed to its value derived from the cosmic microwave background radiation (CMB).

In this paper we have examined the validity of the CET definition of gravitational entropy in the context of some exact cosmological solutions of the Einstein's field equations. In the next section we will first discuss the formalism of calculating gravitational energy density and temperature, which then provides us with the formalism of defining the gravitational entropy of a physical system. We begin Section III by describing the covariant 1 + 3 splitting of spacetime, which we have used to analyse the behaviour of some cosmological models with regard to the Weyl curvature. We have calculated all the relevant functions like the normalized epoch function, gravitational energy density, gravitational temperature and the gravitational entropy for these models. The discussions and conclusions are presented in Section IV.

\section{Gravitational Entropy}

In the following proposals, the measure of gravitational entropy is based on the Bel-Robinson tensor, which is defined in terms of the Weyl tensor in the form \cite{bel,robinson}

\begin{equation}
T_{abcd}\equiv \frac{1}{4}\left(C_{eabf}C^{e\; \; \; \; f}_{\; \; c d \; \;}+C^{*}_{\; e a b f}C^{* \; e \; \; \; \; f}_{\; \; \; \; \; c d \; \;}\right),
\end{equation}
where $ C^{*}_{abcd}=\frac{1}{2}\eta_{abef}C^{ef}_{\phantom{ef}cd} $ is the dual of the Weyl tensor.

The important property of this tensor is that it is overall symmetric, tracefree, and is covariantly conserved in vacuum or in presence of the cosmological constant. The factor of $1/4$ gives a natural interpretation of the Bel-Robinson tensor in terms of the Weyl spinor \cite{PR}. A measure of gravitational entropy was constructed by Pelavas and Lake \cite{PL} and Pelavas and Coley \cite{PC}, which had the form
\begin{equation}
S=\int W d\tau,
\end{equation}
where the epoch function $ W $ was defined using the Bel-Robinson tensor $T_{abcd}$ and the observer four velocity. The epoch function so constructed was therefore observer dependent and non-negative.

Exploring the correspondence between electromagnetism and general relativity, Maartens and Basset \cite{MB} considered a $1 + 3$ covariant, nonperturbative approach, where the free gravitational field was covariantly characterized by the Weyl gravito-electric and gravito-magnetic spatial tensor fields. They demonstrated the covariant analogy between the tensor Bianchi equations and the vector Maxwell equations, and presented the important result that the Bel-Robinson (BR) tensor is a ``unique Maxwellian tensor'' which could be constructed from the Weyl tensor, which behaves as the ``super energy-momentum'' tensor for the gravitational fields. The only problem was that the dimension of the BR tensor is $L^{-4}$ and not $L^{-2}$ (where $L$ is the unit of length), which is the expected dimension for the energy momentum tensor. Based on this work, Clifton, Ellis and Tavakol \cite{CET} proposed that the symmetric $2$-index square root $t_{ab}$, of the BR tensor should act as the effective energy momentum tensor for free gravitational field. Subsequently Goswami and Ellis \cite{GE} constructed a tensor describing the interaction between free gravity and matter, which is taken to be the symmetric two index square root of the BR tensor. Since we intend to examine the CET proposal of gravitational entropy as applied to a few interesting cosmological models, so we will now present a brief review of the CET proposal.

\subsection{The CET proposal}
In the CET proposal of gravitational entropy \cite{CET}, rather than constructing the entropy measure as an integral along a timelike curve, they employed integrals over spacelike hypersurfaces. In section 2 of their paper they laid down a list of requirements for a viable measure of gravitational entropy, $S_{grav}$. They used the gravito-electromagnetic properties of the Weyl tensor and the 1+3 decomposition of the equations, to express the epoch function as follows:
\begin{equation}
W=T_{abcd}u^{a}u^{b}u^{c}u^{d}=\dfrac{1}{4}\left(E_{a}^{b}E_{b}^{a}+ H_{a}^{b}H_{b}^{a}\right).
\end{equation}

Here $ W $ is the ``Super energy density'' and $ E_{ab},H_{ab} $ are the electric and magnetic parts of the Weyl tensor respectively. The inhomogeneous distribution would require that either $E_{ab}$ or $H_{ab}$ is non-zero, so that $W > 0$, which means that inhomogeneity requires both
anisotropy and a non-zero $W$.

From this symmetric and tracefree four-index tensor $T_{abcd}$, one can define a symmetric two-index “square-root”, $t_{ab}$, which is a solution of the equation
\begin{equation}\label{belrob}
T_{abcd}=t_{(ab}t_{cd)}-\frac{1}{2} t_{e (a} t_{b}^{\phantom{b} e} g_{c d)} - \frac{1}{4} t_{e}^{\phantom{e} e} t_{(a b} g_{cd)} +\frac{1}{24} \left( t_{ef} t^{ef} +\frac{1}{2} (t_{e}^{\phantom{e}e})^2 \right)g_{(ab}g_{cd)}.
\end{equation}
The right hand side of this equation constitutes the only totally symmetric and tracefree four index
tensor, the quadratic $ t_{ab}$, that may be constructed. For any solution, $ t_{ab}$, there exists another solution $\epsilon t_{ab} + f g_{ab}$, where $ \epsilon=\pm 1 $, and $ f $ is an arbitrary function.
In spacetimes of Petrov type D or N, although the solution to the above equation is unique for a tracefree $t_{ab}$, but that does not necessarily lead to a quantity that is conserved in vacuum. Therefore the square-root of the Bel-Robinson may be chosen to inherit its tracefree property, or for its conservation in vacuum, but not necessarily both at the same time. For Petrov type D spacetimes, with two double principal null directions, and a Coulomb-like gravitational field, the tracefree square-root can be written in the form
\begin{equation}\label{tracel}
t_{ab}=3\epsilon|\Psi_{2}|(m_{(a}\bar{m}_{b)}+l_{(a}k_{b)}),
\end{equation}
where $\Psi_2= C_{abcd} k^a m^b \bar{m}^c l^d$ is the only non-zero Weyl scalar. The complex null tetrad is defined as
\begin{equation}
\label{nullt}
m^a = \frac{1}{\sqrt{2}} \left( x^a - i y^a \right), \quad
l^a = \frac{1}{\sqrt{2}} \left( u^a - z^a \right), \quad {\rm and} \quad
k^a = \frac{1}{\sqrt{2}} \left( u^a + z^a \right),
\end{equation}
where $x^a$, $y^a$ and $z^a$ are spacelike unit vectors, which form an orthonormal basis together with $u^a$, $g_{ab} =2 m_{(a} \bar{m}_{b)} - 2 k_{(a} l_{b)}$, with $l^a$ and $k^a$ being aligned with the principal null directions.
The effective energy-momentum, $ \tau_{ab} $, of the Coulomb-like gravitational fields present in a Petrov type D spacetime, was assumed to be given by the solution to \eqref{belrob}, with a traceless part described by \eqref{tracel}, so that
\begin{equation}\label{diff}
8\pi\tau_{ab}=\alpha[3\epsilon\vert\Psi_{2}\vert(m_{(a}\bar{m}_{b)}+l_{(a}k_{b)})+fg_{ab}] \nonumber\\
             =\alpha \left[ \left( \frac{3}{2} \epsilon \vert \Psi_2 \vert +f \right) \left( x_a x_b +y_a y_b\right) - \left( \frac{3}{2} \epsilon \vert \Psi_2 \vert - f \right)\left(  z_a z_b -u^a u^b \right) \right].
\end{equation}
Here $ \alpha $ is an unknown constant which must be determined. By contracting this effective energy-momentum tensor with the timelike unit vector, $ u^{a} $, and the projection tensor, $ h_{ab} $, one obtains the effective energy density, pressure and momentum density, which are given below:
\begin{equation}\label{var}
8\pi \rho_{\rm grav} = \alpha \left( \frac{3}{2} \epsilon \vert \Psi_2 \vert -f \right), \quad
8\pi p_{\rm grav} = \alpha\left( \frac{1}{2} \epsilon \vert \Psi_2 \vert +f \right), \quad
8\pi \pi^{\rm grav}_{ab} = \frac{\alpha}{2} \epsilon \vert \Psi_2 \vert  \left( x_a x_b +y_a y_b -z_a z_b + u^a u^b \right),
\end{equation}
and $q^{\rm grav}_a = 0$. These quantities seem to obey the equations that are closely analogous to those of matter fields, and therefore these equations were used to construct a definition of gravitational entropy. The free function $ f $ can be removed by imposing the energy conservation condition in vacuum for the effective energy momentum tensor, and simultaneously (\ref{belrob}) is to be imposed to get the functional form of $ f $. To do so, equation \eqref{diff} is to be differentiated, and at the same time one has to use the relation
\begin{equation}\label{psi2}
|\Psi_{2}|=\sqrt{\dfrac{2W}{3}},
\end{equation}
and equation $ (45) $ of \cite{MB}, which represents the covariant non-perturbative generalization of Bel's linearized conservation equation in \cite{bel}. Finally one can derive the functional form of $ f $ as follows:
\begin{equation}
f=-\frac{1}{2} \epsilon \vert \Psi_2 \vert +\lambda_1,
\end{equation}
where $ \lambda_1 $ is an arbitrary constant which may be set to zero, as it does not affect the relevant thermodynamic quantities. Consequently the effective energy-momentum tensor is obtained in the form
\begin{equation}
8 \pi \tau_{ab} =  \epsilon \alpha \sqrt{\frac{2 W}{3}} \left( x_a x_b +y_a y_b\right) - 2\epsilon \alpha \sqrt{\frac{2 W}{3}}  \left(  z_a z_b -u^a u^b \right),
\end{equation}
and the effective energy density and the pressure in the free gravitational field are obtained as
\begin{equation}
8\pi\rho_{grav}=2\alpha\sqrt{\dfrac{2W}{3}},  \qquad  {\rm and} \qquad
p_{\rm grav} =0,
\end{equation}
where $\epsilon =+1$, so that $\rho_{\rm grav} \geq 0$. The anisotropic pressure and momentum density remain unaffected according to equation (\ref{var}) and the relation $q^{\rm grav}_a=0$, because the function $f$ occurs only in the trace of $\tau_{ab}$. These relations constitute the ``unique expressions'' for the effective energy density and pressure of free gravitational field in Petrov type D spacetimes \cite{CET}, which are determined from the square-root of the Bel-Robinson tensor $ T_{abcd} $ by imposing the condition of energy conservation in vacuum, i.e., $u_a \tau^{ab}_{\phantom{ab};b}=0$, and the positivity of energy density. Finally the  relation (\ref{psi2}) implies that many properties of the Bel-Robinson tensor are inherited by the effective energy-momentum tensor $ \tau_{ab} $.
In order to calculate the entropy according to thermodynamic prescriptions, one must know the {``temperature'', $T_{\rm grav}$}, of the free gravitational fields. For that purpose, one must have possess some knowledge about the underlying microscopic theory. Naturally, the CET proposal assumed that a thermodynamic treatment of the free gravitational field is very much similar to that of standard thermodynamics. This was the motivation for looking into the results of black hole thermodynamics, and quantum field theory in curved spacetimes. Therefore they required the definition of temperature to be local (instead of being defined for horizons only), which reproduced the expected results from semi-classical calculations in Schwarzschild and de Sitter spacetimes. The temperature at any point in spacetime was given by the following expression:
\begin{equation}
T_{grav}=\dfrac{|u_{a;b}l^{a}k^{b}|}{\pi}=\dfrac{|\dot{u_{a}}z^{a}+H+\sigma_{ab}z^{a}z^{b}|}{2\pi},
\end{equation}
where $z^{a}$ is a spacelike unit vector aligned with the Weyl principal tetrad, and  $ H=\dfrac{\Theta}{3} $ is the isotropic Hubble rate.

\section{Gravitational entropy of some cosmological models}

From earlier studies \cite{CET} it is known that the CET proposal of gravitational entropy is applicable only to Petrov type D spacetimes in four dimensional GR. Therefore, in the following sequel we will explore some Petrov type D spacetimes representing the various phases of evolution of the universe filled with ideal irrotational fluids.

We will use the covariant $1+3$ splitting of spacetime \cite{Elst1,Elst2,Elst3} with the timelike vector field $ u^{a} $ and the projection tensor $ h_{ab} $ satisfying the following relations
\begin{equation}
  U^{a}\,_{b} \coloneqq u^{a}u_{b},\;\;\;\;\;   g_{ab} \coloneqq -u_{a}u_{b}+h_{ab},
\end{equation}
\begin{equation}
 U^{a}\,_{c} U^{c}\,_{b}=  U^{a}\,_{b},\;\;\;\;\;  U^{a}\,_{b}u^{b}= u^{a},\;\;\;\;\;  U^{a}\,_{a} =-1,
\end{equation}
\begin{equation}
 h^{a}\,_{c} h^{c}\,_{b}= h^{a}\,_{b},\;\;\;\;\;  h^{a}\,_{b}u^{b}=0,\;\;\;\;\; h^{a}\,_{a}=3.
\end{equation}
The covariant time derivative (represented by a dot) and the projected spatial derivative (represented by $\mathbf{D})$ using the projection tensor $ h_{ab} $ are given by:
\begin{equation}
\dot{A}^{a...}_{b...}=u^{c}\bigtriangledown_{c}A^{a...}_{b...}, \;\;\;\;\;    \mathbf{D}_{a}A^{b...}_{c...} \equiv h_{a}^{p}h^{b}_{q}...h_{c}^{r} \mathbf{\bigtriangledown}_{p}A^{q...}_{r...},
\end{equation}

\begin{equation}
\dot{U}^{<ab>}=\dot{h}^{<ab>}=0,\;\;\;\;\; \mathbf{D}_{a}U_{bc} = \mathbf{D}_{a}h_{bc}=0\,.
\end{equation}
Here the angular brackets `$ <> $' denote the symmetric and trace-free part of a tensor.
The $ 3-$volume element is defined as:
\begin{equation}
\epsilon \coloneqq -\epsilon_{defg}h^{d}\,_{a}h^{e}\,_{b}h^{f}\,_{c}u^{g}=u^{g}\epsilon_{gdef}h^{d}\,_{a}h^{e}\,_{b}h^{f}\,_{c}\,.
\end{equation}

The kinematical variables are obtained from the covariant derivative of $ \mathbf{u}$, which is given by
\begin{equation}
\nabla_{a}u_{b}=-u_{a}\dot{u}_{b}+\mathbf{D}_{a}u_{b}\coloneqq -u_{a}\dot{u}_{b}+\frac{1}{3}\Theta h_{ab}+\sigma_{ab}+\epsilon_{abc}\omega^{c},
\end{equation}
where the kinematical variables are defined by
\begin{equation}
\dot{u}^{a} \coloneqq u^{b}\nabla_{b}u^{a},\;\;\;  \Theta \coloneqq \mathbf{D}_{a}u^{a},\;\;\;  \sigma_{ab} \coloneqq \mathbf{D}_{<a}u_{b>},\;\; \;  \omega^{a}\coloneqq \epsilon^{abc} \mathbf{D}_{b}u_{c}\; .
\end{equation}
The matter variables are defined as the following:
\begin{equation}
\mu \coloneqq T_{ab}u^{a}u^{b},\;\;\;  q^{a}  \coloneqq -T_{cb}h^{ca}u^{b},\;\;\;  p \coloneqq \dfrac{1}{3} T_{ab}h^{ab},\;\;\;  \pi_{ab} \coloneqq T_{cd} h^{c}\,_{<a}h^{d}\,_{b>}\;.
\end{equation}

The electric and magnetic parts of the Weyl curvature tensor are given by
\begin{equation}
E_{ab} \coloneqq C_{cdef}h^{c}\,_{a}u^{d}h^{e}\,_{b}u^{f},\;\;\;\;\;  H_{ab} \coloneqq (-\frac{1}{2}\epsilon_{cdgh}C^{gh}\,_{ef})h^{c}\,_{a}u^{d}h^{e}\,_{b}u^{f}\;.
\end{equation}

In order to drive in to the point that the gravitational entropy measured in terms of the Weyl tensor indeed fulfills the physical requirement that it reflects the inherent anisotropy of the spacetime, we will take into consideration two equations.

One of the Ricci identities is given by \cite{Elst1}:
\begin{equation}\label{d1}
\dot{\sigma}^{<ab>}-\mathbf{D}^{<a}\dot{u}^{b>}=-\dfrac{2}{3}\Theta\sigma^{ab}+\dot{u}^{<a}\dot{u}^{b>}-\sigma^{<a}\,_{c}\,\sigma^{b>c}-\omega^{<a} \omega^{b>}-\left(E^{ab}-\frac{1}{2}\pi^{ab}\right),
\end{equation}
and also one of the contracted Bianchi identities \cite{Elst1} is
\begin{align}\label{d2}
\left(\dot{E}^{<ab>}+\frac{1}{2}\dot{\pi}^{<ab>}\right)-\epsilon^{cd<a}\mathbf{D}_{c}H^{b>}\,_{d}+\frac{1}{2}\mathbf{D}^{<a}q^{b>}
=-\frac{1}{2}(\mu+p)\sigma^{ab}-\Theta\left(E^{ab}+\dfrac{1}{6}\pi^{ab}\right) \nonumber\\
+3\sigma^{<a}\,_{c}\left(E^{b>c}-\frac{1}{6}\pi^{b>c}\right)-\dot{u}^{<a}q^{b>}
\nonumber\\
+\epsilon^{cd<a}\left[2\dot{u}_{c}H^{b>}\,_{d}+\omega_{c}\left(E^{b>}\,_{d}+\frac{1}{2}\pi^{b>}\,_{d}\right)\right]
\end{align}
A very important point is to be noted in this context, i.e., equations (\ref{d1}) and (\ref{d2}) together define a two way relationship between the shear and the electric part of Weyl tensor. The electric Weyl drives the evolution of shear and matter density, together with the shear which drives the evolution of the electric Weyl. From these two equations we can clearly identify the physical processes behind the generation of gravitational entropy, i.e. it is indeed generated from the anisotropies of the universe.

\subsection{FLRW model}
Let us begin with the most common and universally accepted cosmological model, the FLRW metric. Since it represents a homogeneous and isotropic universe, the gravitational entropy function is expected to vanish in this case. However, that does not violate the WCH, since gravitational entropy is associated with inhomogeneity.

Here we are only considering the zero spatial curvature case ($k=0$) but the results are valid for any of the three cases: $k=0,\ \pm1$. The generality of the property of conformal flatness for any spatial curvature in FLRW models has been shown explicitly in \cite{frw1}. The flat FLRW metric is given by
\begin{equation}
ds^2=-dt^2 + A^2(t)(dx^2+dy^2+dz^2),
\end{equation}
where $ A(t) $ is the scale factor. This is a conformally flat algebraically special metric, i.e., the Weyl curvature is zero in this case, thereby making it a Petrov type O spacetime, which is a subclass of the algebraically more general Petrov type D spacetime. Consequently all the Weyl curvature components are zero: $ \Psi_{0}=..=\Psi_{4}=0 $. For the sake of completeness let us consider the following four vectors in conformity with the Weyl principal tetrad:
\begin{equation}
u^{a}=\left(1,0,0,0\right),
\end{equation}
\begin{equation}
z^{a}=\left(0,\frac{1}{A},0,0\right),
\end{equation}
where $ u^{a} $ and $ z^{a} $ forms the null cone, and the $ (m,\bar{m}) $ plane is covered by following two four vectors:
\begin{equation}
x^{a}=\left(0,0,\frac{1}{A},0\right),
\end{equation}
\begin{equation}
y^{a}=\left(0,0,0,\frac{1}{A}\right).
\end{equation}
Using our chosen $ u^{a} $, we can determine the expansion scalar $ \Theta $ along with other quantities like the acceleration, shear tensor and rotation tensor. The expression for the expansion is obvious and matches with the isotropic three dimensional volume expansion:
\begin{equation}\label{expfrw}
\Theta=\frac{3\dot{A}}{A}.
\end{equation}
Subsequently all the components of the shear tensor, $ \sigma_{ab} $, are zero along with vanishing acceleration, and rotation tensor. We thus obtain
\begin{equation}
\rho_{grav}=\frac{\alpha}{4\pi}\vert\Psi_{2}\vert=0,
\end{equation}
\begin{equation}\label{Tfrw}
T_{grav}=\left\vert\dfrac{\dot{u_{a}}z^{a}+H+\sigma_{ab}z^{a}z^{b}}{2\pi}\right\vert=\frac{1}{2\pi}\left\vert\frac{\dot{A}}{A}\right\vert,
\end{equation}
and
\begin{equation}
S_{grav}=\int_{V}\dfrac{\rho_{grav}v}{T_{grav}}=0,
\end{equation}
which should be obvious because the metric is conformally flat, and the measure of free gravitational energy density depends on the Bel–Robinson tensor which is constructed out of the Weyl tensor and its dual \cite{CET}. The purpose of this rather simple exercise is to show that the gravitational temperature is nonzero and finite in spite of the fact that its free gravitational energy density vanishes. From the expression of the gravitational temperature \eqref{Tfrw}, it is clear that this quantity depends on the expansion of the FLRW spacetime, as given in \eqref{expfrw}. To put this argument in the proper context, let us consider different epochs of evolution of the universe. In the radiation dominated era, we know that the scale factor $A(t)$ varies as $ \sqrt{t} $, yielding the gravitational temperature as $ T_{grav}=\frac{1}{2\pi}\left(\frac{1}{2t} \right)$. Similarly for the matter dominated era we can take the scale factor to be $ A(t)\varpropto t^{2/3} $, so that $ T_{grav}=\frac{1}{3\pi}\left(\frac{1}{t} \right)$. Note that in both the radiation and matter dominated eras, the gravitational temperature varies inversely with time. Lastly, for the dark energy dominated era, $ A(t)\varpropto \exp(H_{0}t) $, where $H_{0} $ is the Hubble constant. Consequently, the gravitational temperature for the dark energy dominated era is $ T_{grav}=\frac{H_{0}}{2\pi}=\frac{1}{2\pi}\sqrt{\frac{\Lambda}{3}} $, which turns out to be directly related to the cosmological constant, $ \Lambda $. Analyzing the expression of gravitational temperature from the radiation dominated era through the matter dominated era into the dark energy dominated era, we can say that the gravitational temperature decreases in course of time and ultimately tends to a value determined by the cosmological constant. Although the gravitational entropy carried by the free gravitational field is zero in these cases, which is in agreement with our understanding of \cite{CET}, but the difference in the gravitational temperature among different cosmological eras hints towards a hidden physics behind it. It is a matter of independent investigation, which we leave aside for the moment.

\subsection{LRS Bianchi I model}
If we break the isotropy, then the next step would be to consider the Bianchi class of spacetimes, namely the most general one: Bianchi type I. This spacetime is algebraically general and doesn't help us to apply the CET proposal. Therefore a more physically interesting case would be to consider the Locally Rotationally Symmetric (LRS) Bianchi type I spacetimes \cite{lrsbi1}, where two of the spatial directions have the same directional scale factor, whereas the third one has a different scale factor. The FLRW metric can be considered to be a special case of this spatially homogeneous and anisotropic model. A general form of the LRS Bianchi I metric is the following \cite{lrsbi2}:
\begin{equation}
ds^2=-dt^2 + A^2(t)dx^2+B^2(t)(dy^2+dz^2).
\end{equation}
Imposing the LRS restriction makes the Bianchi I spacetime algebraically special, with Petrov D classification. Therefore we will have one nonzero weyl component $ \Psi_{2} $ which will give us the free gravitational energy density. Choosing our vectors in accordance with the Weyl principal tetrad, we can calculate the necessary variables as given below. The expression for the expansion scalar and the components of the shear tensor are the following:
\begin{equation}
\Theta=\left(\dfrac{\dot{A}}{A}+\dfrac{2\dot{B}}{B}\right),
\end{equation}
\begin{equation}
\sigma_{xx}=-\left(\dfrac{2A(\dot{B}A-\dot{A}B)}{3B}\right), \qquad \sigma_{yy}=\dfrac{B(\dot{B}A-\dot{A}B)}{3A}, \qquad \sigma_{zz}=\dfrac{B(\dot{B}A-\dot{A}B)}{3A}.
\end{equation}
Using these components, the shear scalar $\sigma^2 $ can be evaluated as follows:
\begin{equation}
\sigma^2=\frac{1}{3}\left(\frac{\dot{B}}{B}-\frac{\dot{A}}{A}\right)^2.
\end{equation}
Consequently the expansion anisotropy is
\begin{equation}
\frac{\sigma^2}{\Theta^2}=\frac{1}{3}\left(\dfrac{\dot{B}A-\dot{A}B}{2A\dot{B}+B\dot{A}}\right)^2.
\end{equation}
The gravitational epoch function can be obtained using the Bel-Robinson tensor, which is given by
\begin{equation}
W=T_{abcd}u^{a}u^{b}u^{c}u^{d}=\dfrac{(B\ddot{B}A-\dot{B}^2A+\dot{A}B\dot{B}-\ddot{A}B^2)^2}{24A^2B^4}.
\end{equation}
For our calculations we have used the following four-vectors:
\begin{equation}
u^{a}=\left(1,0,0,0\right),
\end{equation}
\begin{equation}
z^{a}=\left(0,\frac{1}{A},0,0\right).
\end{equation}
Here $u^{a} \, \textrm{and} \, z^{a}  $ form the null vectors $ l^{a} $ and $ k^{a} $ generating the null cone. The vectors
\begin{equation}
x^{a}=\left(0,0,\frac{1}{A},0\right),
\end{equation}
and
\begin{equation}
y^{a}=\left(0,0,0,\frac{1}{A}\right),
\end{equation}
form the $ (m,\bar{m}) $ plane. Using these values, we calculate the free gravitational energy density $\rho_{grav}  $, which is found to be
\begin{equation}
\rho_{grav}=\frac{\alpha}{4\pi} \left\vert\dfrac{B\ddot{B}A-\dot{B}^2A+\dot{A}B\dot{B}-\ddot{A}B^2}{6B^2A}\right\vert.
\end{equation}
We can use our null tetrad to determine the Weyl-NP scalars. As it is of Petrov class D, the only nonzero component is $\Psi_{2}  $. The values of $ \Psi_{i} $ are the following: $ \Psi_{0}=0, \Psi_{1}=0, \Psi_{2}=\dfrac{(B\ddot{B}A-\dot{B}^2A+\dot{A}B\dot{B}-\ddot{A}B^2)}{6B^2A}, \Psi_{3}=0 $ and $ \Psi_{4}=0 $. Therefore the relation $ |\Psi_{2}|=\sqrt{\dfrac{2W}{3}} $ is satisfied here, indicating that the choice of this nonholonomic tetrad is in accordance with the local light cone structure. Subsequently the gravitational temperature is given by
\begin{equation}
T_{grav}=\frac{1}{2\pi}\left\vert\frac{\dot{A}}{A}\right\vert .
\end{equation}
Surprisingly, the contribution from the shear and the expansion oppose each other, and the only contributing component remains when the isotropy is broken. The anisotropy in the geometry gives rise to this opposing feature between the expansion and the shear, giving us a net contribution to the temperature, meaning that the gravitational temperature is powered by the anisotropy of a homogeneous system. In order to have a physically viable universe, the directional Hubble parameter $\left\vert\frac{\dot{A}}{A}\right\vert  $ should decrease with time $ t $. Finally the gravitational entropy is given by the expression:
\begin{equation}\label{enbi0}
S_{grav}=\int_{V}\dfrac{\rho_{grav}v}{T_{grav}}=\frac{\alpha}{12}\left\vert\dfrac{A(B\ddot{B}A-\dot{B}^2A+\dot{A}B\dot{B} -\ddot{A}B^2)}{\dot{A}}\right\vert\int_{V}dxdydz .
\end{equation}
In order to analyze the above expression of gravitational entropy, we notice that
\begin{equation}\label{wbi}
C_{abcd}C^{abcd}=\dfrac{4(\dot{B}^2A-\dot{A}B\dot{B}-B\ddot{B}A+\ddot{A}B^2)^2}{3A^2B^4} .
\end{equation}
Now let us impose a strong condition like the one which Penrose originally proposed, i.e., the Weyl scalar should increase with time as the universe expands. In the denominator of the Weyl curvature scalar we have the volume $ V (=AB^2) $ squared, and in order that the Weyl curvature scalar may increase, the numerator must dominate and it should increase faster than the denominator. Therefore $ \left\vert\dfrac{A(B\ddot{B}A-\dot{B}^2A+\dot{A}B\dot{B}-\ddot{A}B^2)}{\dot{A}}\right\vert $ is increasing monotonically with time as $\left\vert\frac{A}{\dot{A}}\right\vert$ is also increasing with time along with the term $ \left\vert(B\ddot{B}A-\dot{B}^2A+\dot{A}B\dot{B}-\ddot{A}B^2)\right\vert $. Therefore it is clear that the expression (\ref{enbi0}) for the gravitational entropy is non negative, and increases monotonically with time.
The above equation (\ref{enbi0}) can be rearranged as follows:
\begin{equation}\label{enbi}
S_{grav}=\dfrac{\alpha}{12}\left\vert \frac{V}{\dot{A}}\left(\sqrt{3}\dot{\sigma}A-\Theta+3\dot{A}\left(\frac{\dot{B}}{B}\right)\right) \right\vert V ,
\end{equation}
where we have taken $ V=AB^2 $. From the equation (\ref{enbi}) it is clear that the gravitational entropy depends on the shear $ \sigma $, and as the rate of shear increases, it contributes positively to the gravitational entropy. Therefore in order to have a physically relevant universe, the gravitational entropy should increase monotonically with time, i.e.
\begin{equation}\label{condition_0}
\dot{S}_{grav}>0,
\end{equation}
which means that
\begin{equation}\label{condition}
\dfrac{d}{dt}\left[\dfrac{A}{\dot{A}}(B\ddot{B}A-\dot{B}^2A+\dot{A}B\dot{B}-\ddot{A}B^2)AB^2\right]>0.
\end{equation}
Now if we choose the condition $ -\beta>0 $, where $ \beta=(\dot{B}^2A-\dot{A}B\dot{B}-B\ddot{B}A+\ddot{A}B^2) $, the above condition \eqref{condition} reduces to the following identity:
\begin{equation}\label{betacond_bi}
\dfrac{\dot{\beta}}{\beta}>\dfrac{\ddot{A}}{\dot{A}}-\left(\dfrac{\dot{A}}{A}+\Theta\right).
\end{equation}
To put the entire analysis in the proper perspective, let us observe the rate of change of Weyl scalar:
\begin{equation}\label{wtbi}
\dfrac{d}{dt}(C_{abcd}C^{abcd})=\dfrac{8}{3A^2B^4}(\beta\dot{\beta}-\beta^2 \Theta).
\end{equation}
From the above condition \eqref{wtbi} the condition of monotonicitally increasing Weyl curvature scalar can be obtained as the following:
\begin{equation}
\dfrac{\dot{\beta}}{\beta}>\Theta, \,\, \textrm{if}  \,\, \beta>0; \qquad \textrm{and} \qquad
\,\,\,\dfrac{\dot{\beta}}{\beta}<\Theta, \,\, \textrm{if}  \,\, \beta<0.
\end{equation}

Let us fix our condition as $ \beta<0 $ for the following analysis.
Note that to obtain a monotonically increasing $ C_{abcd}C^{abcd} $ we need to satisfy the condition $\dfrac{\dot{\beta}}{\beta}<\Theta$, which is not always true when we want to have monotonically increasing gravitational entropy. Now to determine the restriction on $ \dfrac{d}{dt}(C_{abcd}C^{abcd}) $ for monotonically increasing gravitational entropy, i.e., $ \dot{S}_{grav}>0 $, once again we consider the identity \eqref{betacond_bi}.
We know that
\begin{equation}\label{monC}
\dfrac{d}{dt}(C_{abcd}C^{abcd})=\dfrac{8}{3A^2B^4}(\beta\dot{\beta}-\beta^2 \Theta)=\dfrac{8}{3A^2B^4}\beta^2\left(\dfrac{\dot{\beta}}{\beta}-\Theta\right).
\end{equation}
Imposing the condition of monotonically increasing gravitational entropy, i.e., \eqref{betacond_bi} on \eqref{monC} we get the following condition :
\begin{equation}
\dfrac{d}{dt}(C_{abcd}C^{abcd})>\dfrac{8}{3A^2B^4}\beta^2\left(\dfrac{\ddot{A}}{\dot{A}}-\dfrac{\dot{A}}{A}-2\Theta\right).
\end{equation}
Now if $\left(\dfrac{\ddot{A}}{\dot{A}}-\dfrac{\dot{A}}{A}-2\Theta\right)>0 , $ then $d/dt({C_{abcd}C^{abcd}})$ is always positive, i.e., the Weyl curvature scalar is monotonically increasing at all times. But if $\left(\dfrac{\ddot{A}}{\dot{A}}-\dfrac{\dot{A}}{A}-2\Theta\right)<0 ,$ then $d/dt(C_{abcd}C^{abcd})$ can be negative, implying that the Weyl curvature may decrease while the gravitational entropy is increasing, which is similar to the situation illustrated by the authors in \cite{greg}. Thus we can conclude that the LRS Bianchi I spacetime with different kinds of matter as their source, must satisfy the above condition \eqref{condition} for the monotonic increase of gravitational entropy. We are keeping this analysis general as it is clear that the validity of condition \eqref{condition} depends on the nature of the source. In short, in order to have a monotonically increasing gravitational entropy, the LRS Bianchi I spacetimes with various matter sources must satisfy the condition \eqref{condition}, or in other words, it must have an increasing Weyl curvature scalar for the condition $\left( \dfrac{\ddot{A}}{\dot{A}} - \dfrac{\dot{A}}{A} - 2\Theta \right)>0 $. The same analysis can be done by assuming $ \beta $ to be positive. Summing up, first we have shown that, if the Weyl curvature is diverging at the initial singularity or is decreasing with increasing time, then the LRS Bianchi I spacetime can have decreasing gravitational entropy thereby violating the Weyl curvature hypothesis, but in the later part of our analysis we also demonstrated that if we need a monotonically increasing gravitational entropy, then depending on the conditions we may either have an increasing or a decreasing Weyl scalar.

\subsection{Liang model}
Let us now consider a spacetime representing the early phase of evolution of the universe to see whether the CET gravitational entropy proposal holds good in this era.
An example of an exact solution of the Einstein's field equation with an irrotational fluid source with the equation of state $p=1/3\mu$, energy density $ \mu=T_{ab}u^{a}u^{b}=3/(4t^2A^2)$, and fluid velocity $u=A^{-1/2}\dfrac{\partial}{\partial t}$, representing the radiation dominated universe, and whose initial singularity is `Friedmann-like' as considered by Liang (1972) \cite{Liang}, is given by the metric
\begin{equation}
ds^2=-Adt^2 + t[A^{-1}dx^2+A^2b^{-2}(dy^2+f^2dz^2)],
\end{equation}
where $ A=1-(4\tilde{\epsilon} b^2 t)/9 $, $b\equiv$ constant, $ f(y)= \sin y , \tilde{\epsilon}=+1$ and $f(y)= \sinh y ,\tilde{\epsilon}=-1$.
In our subsequent calculations we will assume $ \tilde{\epsilon}=1 $.

The expansion scalar obtained in this model is
\begin{equation}
\Theta=\dfrac{9}{2t}\dfrac{(9-8b^{2}t)}{(9-4b^{2}t)^{3/2}}.
\end{equation}
Apparently this model expands in the interval $ 0<t<\dfrac{9}{8b^{2}} $, since $ \Theta>0 $ in this range, and then shrinks in the interval $\dfrac{9}{8b^{2}}<t<\dfrac{9}{4b^{2}}$, for which $ \Theta<0 $ . Therefore we will consider the range of $ t $ as $ 0<t<\dfrac{9}{8b^{2}} $, since it represents an expanding universe. The acceleration vector and the vorticity tensor turns out to be zero in this case.

The corresponding components of the shear tensor are
\begin{equation}
\sigma_{xx}=\dfrac{108b^{2}t}{(-4b^{2}t+9)^{5/2}}, \qquad \sigma_{yy}=\dfrac{-2\sqrt{-4b^{2}t+9}t}{27}, \qquad \sigma_{zz}=\dfrac{-2sin^{2}y\sqrt{-4b^{2}t+9}t}{27}.
\end{equation}

Therefore we can evaluate the shear scalar and the expression is given by the following:
\begin{equation}
\sigma^2=\dfrac{108 b^4}{(9-4b^{2}t)^3}.
\end{equation}
From the above expression it is evident that as time $ t $ increases, $\sigma^2 $ i.e. the shear scalar also increases. An important parameter in these models is the ratio $\sigma^2 / \Theta^2$, which is found to be given by
\begin{equation}
\dfrac{\sigma^2}{\Theta^2}=\dfrac{24b^4t}{(9-8b^2t)}.
\end{equation}
It is already known that the ratio of the shear scalar to the expansion scalar (known as \emph{expansion anisotropy}) is a good measure of anisotropy \cite{ans1,ans2}, and we can easily check that the ratio in this case is increasing in the allowed range of time. Thus the universe begins from an isotropic singularity (as the ratio vanishes at $ t=0 $) and then the anisotropy increases with time as the universe expands, thereby fulfilling the requirement of inhomogeneity \cite{CET}.

We now compute the velocity dependent gravitational epoch function for this metric using the Bel-Robinson tensor:
\begin{equation}
W=T_{abcd}u^{a}u^{b}u^{c}u^{d}=\dfrac{2b^{4}}{27A^{6}t^{2}}.
\end{equation}
As we have both the anisotropy and the nonzero $ W $ (corresponding to tidal forces as the magnetic part of the Weyl tensor is zero for Petrov type D spacetimes), it is eligible for the calculation of gravitational entropy, as per the criterion set in \cite{CET}.
The normalized dimensionless scalar constructed from this quantity has the form
\begin{equation}
\tilde{P}=\dfrac{W}{\mu^{2}}=\left(\dfrac{32}{243}\right)\dfrac{b^{4}t^{2}}{A^{2}}.
\end{equation}

As $t\rightarrow 0^{+}$, the normalized Bel-Robinson epoch function vanishes, i.e. $ \tilde{P}\rightarrow 0 $, and $ \tilde{P} $ increases monotonically  as one moves away from the isotropic singularity.

For the sake of computation, we will use the following timelike and spacelike unit vectors in accordance with the Weyl principal tetrad:
\begin{equation}
u^{a}=\left(\dfrac{3}{\sqrt{-4b^{2}t+9}},0,0,0\right),
\end{equation}
and
\begin{equation}
z^{a}=\left(0,\dfrac{1}{3}\sqrt{\dfrac{-4b^2 t+9}{t}},0,0\right).
\end{equation}

The null cone is defined by the vectors $ k^{a}$ and $ l^{a}$ (which therefore lie in the $t,x$ plane). The $ (m,\bar{m}) $ plane is defined by $ m^{a}$, where the spacelike vectors are defined as $ x^{a}=\left(0,0,\dfrac{9b}{\sqrt{t(9-4b^{2}t)^{2}}},0\right) $ and $y^{a}=\left(0,0,0,\dfrac{9b}{\sqrt{t(9-4b^{2}t)^{2}\sin^{2}y}}\right)  $.

The gravitational energy density for this Petrov type D spacetime, obtained from the epoch function $W$, is given by
\begin{equation}
\rho_{grav}=\dfrac{\alpha}{18\pi}\dfrac{b^{2}}{A^{3}t}.
\end{equation}
This spacetime is of Petrov D type and the Weyl scalars are $ \Psi_{0}=0, \Psi_{1}=0, \Psi_{2}=-\dfrac{2b^{2}}{9tA^{3}}, \Psi_{3}=0, \textrm{and} \Psi_{4}=0$. Therefore the relation $ |\Psi_{2}|=\sqrt{\dfrac{2W}{3}} $ is satisfied in this case (as given in eqn.(\ref{psi2})).

The gravitational temperature is given by the expression
\begin{equation}\label{t1}
T_{grav}=\dfrac{1}{8\pi t A^{3/2}}.
\end{equation}
We can see that in order to have a non-negative gravitational energy density and temperature, we require the condition $ A>0 $, which also implies that $0<t<\dfrac{9}{4b^{2}} $. Finally, from the definition of gravitational entropy we have
\begin{equation}\label{s1}
S_{grav}=\int_{V} \dfrac{\rho_{grav}v}{T_{grav}}=\dfrac{4\alpha t^{3/2}}{9}\int_{V} dx \textrm{sin}y dy dz=\dfrac{4\alpha t^{3/2}}{9}\int_{0}^{x} dx \int_{0}^{y}\textrm{sin}y dy \int_{0}^{z}dz.
\end{equation}
Thus
\begin{equation}\label{s1n}
S_{grav}= \dfrac{4\alpha t^{3/2}}{9}x(1-cosy)z .
\end{equation}
Here we can identify $(x,y,z)$ as $(r,\theta,\phi)$, where $x$ acts as the ``radial'' co-ordinate, $\theta \in [0, \pi]$, $\phi \in [0, 2\pi]$ and the $(y,z)$ plane is the $(m,\bar{m})$ plane. Therefore the resulting expression for gravitational entropy will be obtained as

\begin{equation}\label{s1n1}
S_{grav}= \dfrac{16\pi\alpha t^{3/2}r}{9} .
\end{equation}

From the above equations (\ref{s1n}) and (\ref{s1n1}), we can see that the gravitational entropy is non-negative and monotonically increasing, leading to structure formation in the universe \cite{CET}. Further, the increase of shear tensor with time corresponds to the evolution of anisotropy in the universe, which leads to an increase in the above mentioned gravitational entropy.
The above analysis clearly shows us that as $ t\rightarrow 0 $, $ A\rightarrow 1 $, so that both the gravitational energy density $\rho_{grav} $ and the temperature $T_{grav} $ blow up. Consequently, in the limit $ t\rightarrow 0 $, the gravitational entropy $ S_{grav}\rightarrow 0 $, which is in agreement with the Weyl curvature hypothesis.

\subsection{Szekeres model}
We now turn to the spatially inhomogeneous models with irrotational dust as source, i.e., the class $\textbf{II}  $  Szekeres solution of the Einstein's field equations, which is known to be a Petrov type D spacetime. The metric under our consideration is the following \cite{GW1982}:
\begin{equation}
ds^{2}=t^{4}[-dt^{2}+dx^{2}+dy^{2}+(A-\beta_{+}t^{2})^{2}dz^{2}],
\end{equation}

where the function $ A $ is defined as
\begin{equation}
A=a(z)+b(z)x+c(z)y-5\beta_{+}(z)(x^{2}+y^{2}).
\end{equation}

In the class $\textbf{II}  $ Szekeres models, the parameters $ a(z), b(z), c(z), \textrm{and}  \beta_{+} $ are arbitrary smooth functions of $ z $, which gives us the freedom of choosing coordinates. For $ \beta_{+}=0 $, the class $\textbf{II}$ Szekeres solution reduces to FLRW metric. We also observe that if we assume $ a=1,b=0,c=0 $ further, i.e, $A=1$, we get the Cartesian form of FLRW metric directly.
The fluid four velocity is defined as $ u=t^{-2}\dfrac{\partial}{\partial t} $ and the energy density is given by
\begin{equation}
\mu=\dfrac{12}{t^{6}\left(1-\left(\dfrac{\beta_{+}}{A}\right)t^{2}\right)}.
\end{equation}
If the energy density is non-negative, then we need the following conditions to be satisfied: $A>0, (A-\beta_{+}t^{2})>0$. This imposes a bound on $ t $ because $ 0<t<\sqrt{\dfrac{A}{\beta_{+}}}$.  The expansion scalar of the universe is obtained as
\begin{equation}
\Theta=\dfrac{2(3A-4\beta_{+}t^{2})}{t^{3}(A-\beta_{+}t^{2})}.
\end{equation}
Thus this model is expanding throughout the cosmic time since $ \Theta>0 $ in the allowed range of $ t $ due to the fact that $ 0<t<\dfrac{\sqrt{3}}{2}\sqrt{\dfrac{A}{\beta_{+}}} $.

The shear tensor in this case is given by
\begin{equation}
\sigma_{xx}=\dfrac{2\beta_{+}t^{3}}{3(A-\beta_{+}t^{2})}, \qquad \sigma_{yy}=\dfrac{2\beta_{+}t^{3}}{3(A-\beta_{+}t^{2})}, \qquad
\sigma_{zz}=\dfrac{4\beta_{+}t^{3}}{3}(\beta_{+}t^{2}-A).
\end{equation}

The shear scalar is given by
\begin{equation}
\sigma^2=\dfrac{8\beta_{+}^2}{9t^2}.
\end{equation}
The expansion anisotropy in this universe is therefore given by
\begin{equation}
\dfrac{\sigma^2}{\Theta^2}=\dfrac{2\beta_{+}^2t^4(A-\beta_{+}t^2)^2}{9(3A-4\beta_{+}t^2)^2}.
\end{equation}
The above ratio vanishes at $ t=0 $, representing an isotropic initial singularity, and subsequently increases with time. Therefore the anisotropy increases with the evolution and expansion of the universe giving rise to structure formation.

Once again using the fluid $ 4- $velocity, we construct the positive scalar from the Bel-Robinson tensor:
\begin{equation}
W=T_{abcd}u^{a}u^{b}u^{c}u^{d}=\dfrac{6\beta_{+}^{2}}{t^{8}(\beta_{+}t^{2}-A)^{2}}.
\end{equation}
Therefore we get the normalized dimensionless scalar in the form
\begin{equation}
\tilde{P}=\dfrac{W}{\mu^{2}}= \dfrac{t^{4}\beta_{+}^{2}}{24A^{2}}.
\end{equation}
Thus, as $ t \rightarrow 0^{+} $, the normalized Bel-Robinson epoch function vanishes $( \tilde{P}\rightarrow 0 )$.
Let us construct the timelike and spacelike unit vectors in accordance with the Weyl principal tetrad so that $ u_{a}u^{a}=-1 $, $ z_{a}z^{a}=1 $ and $ u_{a}z^{a}=0 $, to get

\begin{equation}
u^{a}=\left(\dfrac{1}{t^{2}},0,0,0\right),
\end{equation}
and
\begin{equation}
z^{a}=\left(0,0,0,\dfrac{1}{t^{2}(A-\beta_{+}t^{2})}\right).
\end{equation}
The $ (m,\bar{m}) $ plane is defined by $ m^{a}$ which is defined in Section II, where the spacelike vectors are now defined as $ x^{a}=\left(0,\dfrac{1}{t^{2}},0,0\right) $ and $y^{a}=\left(0,0,\dfrac{1}{t^{2}},0\right)  $.

From the definition of the gravitational energy density of Petrov type D spacetimes, we get
\begin{equation}
\rho_{grav}=\dfrac{\alpha\beta_{+}}{2\pi t^{4}(A-\beta_{+}t^{2})},
\end{equation}
The above expression of gravitational energy density clearly indicates that for the non-negativity of the gravitational energy density, the following conditions must be fulfilled: $ \beta_{+}>0, A> \beta_{+}t^{2}.$ Therefore $ A $ must be a positive quantity.
The Weyl scalars are: $ \Psi_{0}=0, \Psi_{1}=0, \Psi_{2}=-\dfrac{2\beta_{+}}{t^{4}(\beta_{+}t^{2}-A)}, \Psi_{3}=0, \textrm{and} \Psi_{4}=0$. So the relation $ |\Psi_{2}|=\sqrt{\dfrac{2W}{3}} $ is now satisfied.

The gravitational temperature is given by
\begin{equation}\label{t2_1}
T_{grav}=\dfrac{(A-2\beta_{+}t^{2})}{\pi t^{3}(A-\beta_{+}t^{2})}.
\end{equation}
From the above equation (\ref{t2_1}) it is clear that we require an additional constraint in the form $(A-2\beta_{+}t^{2})>0  $ in order to ensure the non-negativity of the temperature. Thus the allowed range of cosmic time should be $ 0<t<\sqrt{\dfrac{A}{2\beta_{+}}} $.

As before, using the relevant definition, we obtain the expression of gravitational entropy as follows
\begin{equation}\label{s2_1}
S_{grav}=\dfrac{\alpha t^{5}}{2} \int_{0}^{x}\int_{0}^{y}\int_{0}^{z}\left[1+\dfrac{\beta_{+}t^{2}}{(A(x,y,z)-2\beta_{+}(z)t^{2})}\right]\beta_{+}dx dy dz =\dfrac{\alpha t^{5}}{2}T(t),
\end{equation}
where $$ T(t)=\int_{0}^{x}\int_{0}^{y}\int_{0}^{z}\left[1+\dfrac{\beta_{+}t^{2}}{(A(x,y,z)-2\beta_{+}(z)t^{2})}\right]\beta_{+}dx dy dz .$$ The term $ T(t) $ is not directly integrable because of the presence of unknown functions, but the term in the parenthesis is increasing monotonically with $ t $ as the denominator of the second term is decreasing with time and the numerator is directly proportional to $ t^2 $.
Therefore although it is not possible to integrate this equation further, the expression (\ref{s2_1}) of the gravitational entropy is not only non-negative but is also monotonically increasing, thereby satisfying the conditions of structure formation as laid down in \cite{CET}. Moreover, as $ t\rightarrow 0^{+} $, both the gravitational energy density $ \rho_{grav} $ and the temperature $ T_{grav} $ diverge, and as a result the gravitational entropy vanishes, i.e., $ S_{grav}\rightarrow 0 $. Further we know that $ \beta_{+}=0 $ gives us the FLRW metric and indeed the expression of gravitational entropy (\ref{s2_1}) reduces to zero in that case.

\subsection{Bianchi VI$_{h}$ model}
Finally we consider a spacetime which fits a general class of solutions of the Einstein's field equations but simple enough to study a perturbed kind of flat spacetime like the perturbed FLRW spacetime. We will show that the deviation from conformal flatness and isotropy leads us to an inhomogeneous spacetime where gravitational entropy is generated.

Wainwright and Anderson \cite{wainander}, showed that in the Bianchi VI$_h$ class of models, a suitable choice of parameters may help to represent the quasi-isotropic stage beginning at the initial singularity, leading to an isotropic singularity for these spacetimes. By assuming the parameter $\alpha_{c}  $ to be small in that model, one can consider deviations about this flat FLRW model. In the line element in \cite{wainander}, we set $\alpha_{s}=0$ and  $\alpha_{m}=1$, so that the line-element in conformal time coordinates is obtained as follows:
\begin{equation}
ds^{2}=\tau^{4/(3\gamma-2)}(-A^{2(\gamma-1)}d\tau^2+A^{2q_{1}}dx^{2}+A^{2q_{2}}e^{2r[s+(3\gamma-2)]x}dy^{2}+A^{2q_{3}}e^{2r[s-(3\gamma-2)]x}dz^{2}),
\end{equation}
where $ A^{2-\gamma}=1+\alpha_{c}\tau^{2}, \quad q_{1}=\dfrac{\gamma}{2}, \quad q_{2}=\dfrac{2-\gamma+s}{4}, \quad q_{3}=\dfrac{2-\gamma-s}{4}, \quad s^{2}=(3\gamma+2)(2-\gamma) , \quad \textrm{and} \quad r^{2}=\dfrac{(3\gamma+2)\alpha_{c}}{4(2-\gamma)(3\gamma-2)^{2}}. $
The parameter denoted by $ \alpha_{c} $ determines the curvature of the spacelike hypersurfaces orthogonal to $ u=A^{1-\gamma} \tau^{-2/(3\gamma-2)}\dfrac{\partial}{\partial \tau}.$
For $ \alpha_{c}=0 $, we obtain the flat FLRW solution.

In full generality, this metric is of Petrov type I, but the CET gravitational entropy measure only works on the Petrov types D and N. Therefore we will only consider the case for $ \gamma=4/3 $ which reduces the spacetime to Petrov type D.
The resulting Petrov type D metric is given by
\begin{equation}
ds^{2}=\tau^{2}\left(-(\alpha_{c} \tau^{2}+1)d\tau^{2}+(\alpha_{c} \tau^{2}+1)^{2}dx^{2}+(\alpha_{c} \tau^{2}+1)^{2}e^{6\sqrt{\alpha_{c}}x}dy^{2}+\dfrac{1}{(\alpha_{c} \tau^{2}+1)}dz^{2}\right).
\end{equation}

The expansion scalar is given by the following expression
\begin{equation} \label{expn_sclr_bian}
\Theta=\dfrac{3(2\alpha_{c}\tau^{2}+1)}{\tau^{2}(\alpha_{c}\tau^{2}+1)^{3/2}},
\end{equation}
and the shear tensor is
\begin{equation}
\sigma_{xx}=\alpha_{c}\tau^{2}\sqrt{\alpha_{c}\tau^{2}+1}, \qquad \sigma_{yy}=\alpha_{c}\tau^{2}e^{6\sqrt{\alpha_{c}}x}\sqrt{\alpha_{c}\tau^{2}+1}, \qquad
\sigma_{zz}=\dfrac{-2\alpha_{c}\tau^{2}}{(\alpha_{c}\tau^{2}+1)^{5/2}}.
\end{equation}
This model is also expanding with time as the expansion $ \Theta>0 $ for all $ \tau $.
In this case, the shear scalar is given by
\begin{equation}
\sigma^2=\dfrac{3\alpha_{c}^2}{(1+\alpha_{c}\tau^2)^3}.
\end{equation}

Once again, as a measure of expansion anisotropy we compute $ \sigma/\Theta $ which is given by the following expression:
\begin{equation}
\dfrac{\sigma^2}{\Theta^2}=\dfrac{\alpha_{c}^2\tau^4}{3(2\alpha_{c}\tau^2+1)^2}.
\end{equation}
Thus the ratio vanishes at $ \tau=0 $ indicating an initial isotropic singularity as in the previous models. It then increases with time and becomes constant as time tends to infinity. Therefore the time evolution of this universe is such that as time increases, the expansion anisotropy increases from the isotropic initial singularity, and gradually the rate of increase of this ratio decreases and finally it becomes more or less constant in the distant future. It may be mentioned here that exact spatially homogeneous cosmologies in which this ratio is constant, were studied in \cite{CGW}.

We can now easily compute the energy density as
\begin{equation}
\mu=\dfrac{3}{\tau^{4}(\alpha_{c}\tau^{2}+1)^{2}}.
\end{equation}
Next we construct the velocity dependent gravitational epoch function from the Bel-Robinson tensor, which yields
\begin{equation}
W=T_{abcd}u^{a}u^{b}u^{c}u^{d}=\dfrac{6\alpha_{c}^{2}}{\tau^{4}(\alpha_{c} \tau^{2}+1)^{6}}.
\end{equation}
In order to construct a dimensionless scalar from this quantity, we normalize the standard epoch function with the square of $ \mu=T_{ab}u^{a}u^{b} $ to get
\begin{equation}
\tilde{P}=\dfrac{W}{\mu^{2}}= \dfrac{2\alpha_{c}^{2}\tau^{4}}{3(\alpha_{c}\tau^{2}+1)^{2}}.
\end{equation}
As $ \tau \rightarrow 0^{+} $, the normalized Bel-Robinson epoch function vanishes: $ \tilde{P}\rightarrow 0 $. Therefore $ \tilde{P} $ behaves appropriately as the isotropic singularity is approached.

Now, for the analysis of the CET gravitational entropy for this spacetime, we will use the following unit vectors, where $ u^{a} $ is a timelike and $ z^{a} $ is a spacelike unit vector. Here we choose our vectors such that they specify a Weyl principal tetrad:

\begin{equation}
u^{a}=\left(\dfrac{1}{\tau\sqrt{\alpha_{c} \tau^{2}+1}},0,0,0\right),
\end{equation}
and
\begin{equation}
z^{a}=\left(0,0,0,\dfrac{\sqrt{\alpha_{c}\tau^{2}+1}}{\tau}\right).
\end{equation}
We note that this choice of tetrads is also supported by the work of Pelavas and Coley in \cite{PC}. In this case, the $ (m,\bar{m}) $ plane is defined by the spacelike vectors $ x^{a}=\left(0,\dfrac{1}{\tau(\alpha_{c}\tau^{2}+1)},0,0\right) $ and $ y^{a}=\left(0,0,\dfrac{1}{\tau(\alpha_{c}\tau^{2}+1)e^{3\sqrt{\alpha_{c}}x}},0\right) $, with the null cone defined by $l^a$ and $k^a$, as mentioned in Section II, along with the definition of $ m^{a}$.

From the definition of gravitational energy density we now obtain
\begin{equation}\label{r3}
\rho_{grav}=\dfrac{\alpha \alpha_{c}}{2\pi \tau^{2}(\alpha_{c}\tau^{2}+1)^{3}}.
\end{equation}
As this spacetime is a Petrov D spacetime, the Weyl scalars are obtained as $ \Psi_{0}=0, \Psi_{1}=0, \Psi_{2}=\dfrac{2\alpha_{c}}{\tau^{2}(1+\alpha_{c}\tau^{2})^{3}}, \Psi_{3}=0, \textrm{and}  \Psi_{4}=0$. Therefore the relation (\ref{psi2}) is satisfied in this case. Similarly the gravitational temperature can be calculated as
\begin{equation}\label{t3}
T_{grav}=\dfrac{1}{2\pi \tau^{2}(\alpha_{c}\tau^{2}+1)^{3/2}}.
\end{equation}
From the above two expressions of the gravitational energy density (\ref{r3}) and the temperature (\ref{t3}), we can clearly observe that as $ \tau \rightarrow 0^{+} $, both $ \rho_{grav} $ and $ T_{grav} $ diverge near the isotropic singularity.

Now integrating over a volume $ V $ on a hypersurface of constant $ \tau $, we get the expression of gravitational entropy
\begin{equation}\label{t4}
S_{grav}=\alpha \alpha_{c}\tau^{3}\int_{V}e^{3\sqrt{\alpha_{c}}x} dx dy dz=\alpha \alpha_{c}\tau^{3}\int_{0}^{x}e^{3\sqrt{\alpha_{c}}x} dx \int_{0}^{y}dy\int_{0}^{z} dz=\dfrac{\alpha \sqrt{\alpha_{c}}}{3}\tau^{3}(e^{3\sqrt{\alpha_{c}}x}-1)yz = \dfrac{\alpha \sqrt{\alpha_{c}}}{3}V\tau^{3},
\end{equation}
where $ V =(e^{3\sqrt{\alpha_{c}}x}-1)yz$ is a term which depends on the spatial volume, and will monotonically increase with increasing spatial dimensions.
We note that it is not possible to determine the bound of $ \tau $ in (\ref{t4}). However, the final expression indicates that not only the gravitational entropy increases with time $ \tau $, but for any increasing volume (i.e., larger values of $ x,y,z $), the gravitational entropy is increasing monotonically. Therefore, once again we find that $ S_{grav}$ is non-negative and monotonic in nature, and as $ \tau \rightarrow 0^{+} $, the gravitational entropy vanishes, i.e. $S_{grav}\rightarrow 0 $, in accordance with Penrose's Weyl curvature hypothesis. It is also evident that for $ \alpha_{c}\simeq0 $, the spacetime becomes FLRW-like, with vanishing gravitational entropy.

\section{Discussions and Conclusions}

In the very first case for the homogeneous and isotropic FLRW universe, we have shown that the gravitational entropy is zero because the space-time is conformally flat, thereby supporting the Weyl curvature hypothesis, as the free gravitational field in this case does not carry any gravitational energy density while maintaining a finite gravitational temperature. In the LRS Bianchi I case, we have shown explicitly that the gravitational entropy is monotonically increasing with time if its Weyl curvature increases with time, but if there are matter sources in the spacetime which cause the Weyl curvature to decrease in course of time, then the LRS Bianchi I spacetime will have a gravitational entropy which decreases with time, thereby violating the Weyl curvature hypothesis. That is, in order to have a non negative monotonically increasing gravitational entropy, the LRS Bianchi I spacetime must have monotonically increasing Weyl curvature. We are neglecting the cases where $\left\vert \frac{\dot{A}}{A} \right\vert$ is increasing with time as it will give rise to unrealistic situations where the gravitational temperature will increase with time and at the initial singularity it is zero or finite. We have also shown explicitly that if we want to have a non negative monotonically increasing gravitational entropy, depending on the conditions imposed by us, the Weyl curvature scalar can either increase or decrease. In other words, under certain conditions the time derivative of gravitational entropy is positive, i.e., the CET entropy is increasing monotonically.

The above analysis indicates that in the Liang, Szekeres and Bianchi $VI_{h}$ models considered by us, the gravitational entropy goes to zero as we approach the initial singularity and increases monotonically with non-negative value in course of time, thereby fulfilling the necessary requirements to be satisfied by the gravitational entropy in these space-times, in order to ensure structure formation in the universe \cite{CET}. We found that in each of these cases, representing different phases of evolution of the universe, the expansion anisotropy increases as times elapses after the initial isotropic singularity, and it appears that there is a correlation between the expansion anisotropy and the gravitational entropy. We also showed in a general formalism, how the shear tensor is related to the Weyl tensor. Therefore the CET formalism in these cases clearly gives us a well behaved entropy measure which increases as the structure formation progresses, resulting in an increase in anisotropy of these universes. These features make these models physically more realistic for describing the actual evolution of the universe. In all the cases we found that the gravitational entropy vanishes at the initial isotropic singularity. Moreover in each of these cases, the gravitational energy density and the temperature are well-behaved throughout the evolution of the conformal time associated with the metric.

In a recent paper \cite{greg} the authors have illustrated a counterexample involving a class of inhomogeneous universes that are supported by a chameleon massless scalar field and exhibit anisotropic spacetime shearing effects. We will now present a careful review of this work and compare it with our present work.

In \cite{greg}, the shear scalar depends on the Hubble parameter and it is increasing with time. Also the gravitational temperature which they are getting, is independent of time, which can always happen, as in the case of dark energy dominated FLRW universe. In their model, the CET entropy is increasing with time, whereas their gravitational energy density is decreasing with time, which is being compensated by gravitational anisotropic pressure. Finally the authors have shown that the time derivative of gravitational entropy is always positive. The interesting point that the authors in \cite{greg} are making is that the Weyl curvature is decreasing with time despite having increasing CET gravitational entropy, claiming that this violates the Weyl curvature hypothesis.

Here, in this paper, we have considered a variety of different classes of cosmological models and have showed explicitly that for each of them the CET gravitational entropy is increasing. The authors in \cite{greg} stated that the difference between our studies is that their calculation yields time independent gravitational temperature whereas we have found the gravitational temperature to be time dependent. We want to clarify that this is a model dependent phenomenon, and we also have such a case for the dark energy dominated FLRW model where the gravitational temperature is time independent. The important thing to note is that although for the LRS Bianchi I case the Weyl curvature must be increasing in time for a monotonically increasing gravitational entropy, it is not so for the other models considered by us. For the Liang, Szekeres and also for the Bianchi VI$_{h}$ model, although the CET entropy is clearly increasing, the magnitude of $\Psi_{2}$ is decreasing with time, similar to that in \cite{greg}. We want to emphasize that this does not the violate the Weyl curvature hypothesis, as the CET proposal was arrived at after various authors worked with different measures of gravitational entropy, especially the initial case, i.e., where the simple Weyl curvature scalar gives the measure of gravitational entropy. This is precisely the motivation for CET and other measures, as there are plenty of spacetimes where the Weyl curvature is decreasing, giving us a decreasing gravitational entropy if we take the simple Weyl curvature entropy as our measure.

We also want to draw attention to the requirements listed in the original CET paper \cite{CET} for a viable gravitational entropy in the form of E1...E5 at the end of Section 2 in their paper. There is no such requirement where the gravitational entropy should increase with the Weyl curvature, although this was the original proposal of Penrose, but this condition was too strong to be satisfied and subsequently new measures of gravitational entropy were proposed in the form of a suitable function of the Weyl curvature. Therefore we must be very clear when we say that it is satisfying the Weyl curvature hypothesis. What we really mean is that the spacetime has a viable non negative gravitational entropy which is monotonically increasing with time, which is also the case for \cite{greg}. Therefore our conclusions are in accordance with each other, i.e., both these papers are getting non negative and monotonically increasing CET gravitational entropy. The surprising result in \cite{greg} is that, throughout the entire evolution, the matter curvature dominates over the Weyl curvature unlike our cases, but in spite of that the CET gravitational entropy is non negative and monotonically increasing, indicating the robustness of the CET definition of gravitational entropy. The authors in \cite{greg} mention that the shear also plays a very important role, which is true as it is not only affecting the dynamics of the electric part of the Weyl tensor resulting in a change in gravitational energy density, but it is also contained in the gravitational temperature. It is to be noted that \cite{greg} showed that increasing Weyl is NOT necessary but is a sufficient condition for an increasing entropy, and throughout this paper we have worked with this sufficient condition.

Further, we must remember that the CET proposal is independent of any specific definition of gravitational temperature. Therefore, theoretically speaking, different measures of gravitational temperature can be employed. Regarding the role of the definition of gravitational temperature, one can say that it will change the exact results for the gravitational entropy but whether it will affect the monotonicity of the gravitational entropy is a matter of separate investigation. In the definition proposed in \cite{CET}, it depends on the acceleration, expansion, shear tensor and the rotation tensor (if we generalize the definition), capturing all the necessary variables. The time independent or dependent cases may arise from the internal dynamics of these variables, i.e., it is model dependent.

In conclusion, we can clearly state that the definitions of gravitational entropy proposed by Pelavas et al. \cite{PC} and Clifton et al. \cite{CET}, i.e.,  $ \tilde{P} $ and  $ S_{grav} $, are in conformity with the Weyl curvature hypothesis in the case of the models considered by us, and provides a very good description of the gravitational entropy on a local scale. It is to be noted that for a large scale description, one needs to employ the method of averaging, similar to that considered in \cite{cet4}.

\section*{Acknowledgments}
SC is grateful to CSIR, Government of India for providing junior research fellowship. SG gratefully acknowledges IUCAA, India for an associateship and CSIR, Government of India for approving the major research project No. 03(1446)/18/EMR-II. SG is also thankful to the Astrophysics and Cosmology Research Unit, School of Mathematics, Statistics and Computer Sciences, UKZN, South Africa, for supporting her short visit. RG thanks National Research Foundation, South Africa, for research support. We thank the anonymous reviewer for the valuable suggestions.

\end{document}